# Development and analysis of novel mission scenarios based on Atmosphere-Breathing Electric Propulsion (ABEP)

S. Vaidya[1*], C. Traub[2], F. Romano[2], G. H. Herdrich[2], Y.-A. Chan[2], S. Fasoulas[2], P. C. E. Roberts[3], N. H. Crisp[3], S. Edmondson[3], S. J. Haigh[3], B. E. A. Holmes[3], A. Macario-Rojas[3], V. T. A. Oiko[3], K. L. Smith[3], L. A. Sinpetru[3], J. Becedas[4], V. Sulliotti-Linner[4], S. Christensen[5], V. Hanessian[5], T. K. Jensen[5], J. Nielsen[5], M. Bisgaard[5], D. Garcia-Almiñana[6], S. Rodriguez-Donaire[6], M. Suerda[6], M. Garcia-Berenguer[6], D. Kataria[7], R. Villain[8], S. Seminari[8], A. Conte[8], B. Belkouchi[8]

[1] University of Pisa, Italy, Lungarno Antonio Pacinotti, 43, 56126 Pisa PI
[2] Institute of Space Systems (IRS), University of Stuttgart, Pfaffenwaldring 29, 70569 Stuttgart, Germany
[3] The University of Manchester, Oxford Road, Manchester, M13 9PL, UK
[4] Elecnor Deimos Satellite Systems, Calle Francia 9, 13500 Puertollano, Spain
[5] GomSpace AS, Langagervej 6, 9220 Aalborg East, Denmark
[6] UPC-BarcelonaTECH, Carrer de Colom 11, 08222 Terrassa, Barcelona, Spain
[7] Mullard Space Science Laboratory (UCL), Holmbury St. Mary, Dorking, RH5 6NT, United Kingdom
[8] Euroconsult, 86 Boulevard de Sébastopol, 75003 Paris, France
*Corresponding author: s.vaidya@studenti.unipi.it

**Abstract:** Operating satellites in Very Low Earth Orbit (VLEO) benefits the already expanding New Space industry in applications including Earth Observation and beyond. However, long-term operations at such low altitudes require propulsion systems to compensate for the large aerodynamic drag forces. When using conventional propulsion systems, the amount of storable propellant limits the maximum mission lifetime. The latter can be avoided by employing Atmosphere-Breathing Electric Propulsion (ABEP) system, which collects the residual atmospheric particles and uses them as propellant for an electric thruster. Thus, the requirement of on-board propellant storage can ideally be nullified. At the Institute of Space Systems (IRS) of the University of Stuttgart, an intake, and a RF Helicon-based Plasma Thruster (IPT) for ABEP system are developed within the Horizons 2020 funded DISCOVERER project. In order to assess possible future use cases, this paper proposes and analyzes several novel ABEP based mission scenarios. Beginning with technology demonstration mission in VLEO, more complex mission scenarios are derived and discussed in detail. These include, amongst others, orbit maintenance around Mars as well as refuelling and space tug missions. The results show that the ABEP system is not only able to compensate drag for orbit maintenance but also capable of performing orbital maneuvers and collect propellant for applications such as Space Tug and Refuelling. Thus, showing a multitude of different future mission applications.

Abbreviations:  VLEO  – Very Low Earth Orbit
VLMO  – Very Low Mars Orbit
ABEP  – Atmosphere-Breathing Electric Propulsion
EO  – Earth Observation
IPT  – RF Helicon-based Plasma Thruster
FDC  – Full drag compensation

## 1. Introduction

Electric propulsion (EP) devices have paved the path for the 'New Space revolution' in Low Earth Orbits (LEO), leading to a rapidly increasing number of start-ups in the field of Earth Observation (EO), Internet of things (IoT) and broadband services [1]. Many of the applications of focus would thereby experience significant improvements from orbits even closer to the ground than LEO, the so-called Very Low Earth Orbits (VLEO), defined as the entirety of orbits with a mean altitude below 450 km [2]. The advantages of these orbits for each type of application, especially EO, can be referred from [2,3], which are the part of DISCOVERER project. However, satellites orbiting in VLEO are subjected to large aerodynamic drag forces and heating effects due to which propulsion as well as thermal control systems become crucial for their design. Moreover, to maintain an orbit at these altitudes, the spacecraft should constantly generate thrust for a continuous drag compensation. By using state-of-the-art spacecraft propulsion systems (chemical or electric), the amount of stored propellant consequently determines the mission lifetime. This constraint can be eliminated by introducing an Atmosphere-Breathing Electric Propulsion (ABEP) system which collects the residual atmospheric gases and supplies them as propellant to an electric thruster [4]. In this case, the satellite's lifetime, in theory, is only dependent on the durability and reliability of its individual components, especially those which come in direct contact with the atmosphere. Some of the atmospheric species can cause significant damage to the satellite structural elements, payload(s) or the thruster components, which in turn would jeopardize the mission. For instance, around Earth, as VLEO is an atomic oxygen (AO) rich environment, the resistance to oxidation of the exposed components will determine the satellite's lifetime [5]. A schematic of an ABEP-based spacecraft is depicted in Figure 1.

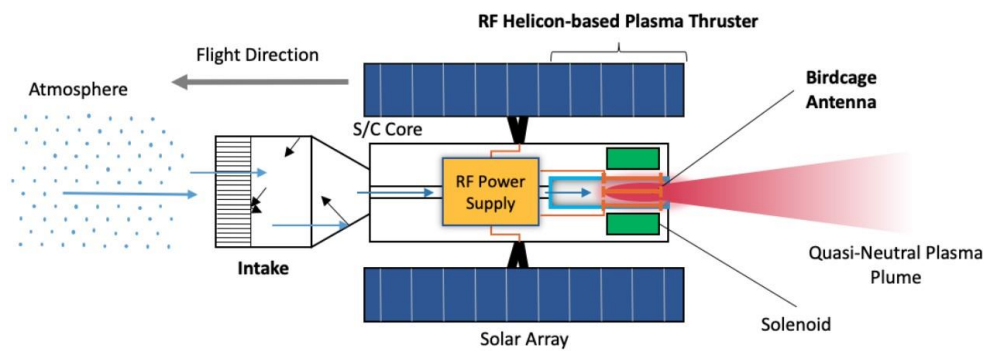

Figure 1: ABEP based spacecraft schematic [4].

Furthermore, this approach is applicable to any celestial body with an atmosphere, provided the thruster and other components employed in the ABEP system are compatible with the atmospheric species encountered around that body. Moreover, it is understood that respective adaptations of relevant subsystems such as e.g., the intake are necessary. This would enable missions such as high-fidelity planetary observation or long term, in-situ atmospheric measurements around Mars, Venus, or Titan [6].

In recent years, several theoretical and experimental investigations are being undertaken in Europe, Japan, the US, and Russia to develop a feasible ABEP technology. In Europe, ESA presented a study evaluating the feasibility of ABEP for a technology demonstration mission on a spacecraft equipped with four Gridded Ion Engines (GIE) [7]. At Sitael, in Italy work is ongoing on development of a concept based on a passive intake and a double-stage EP device with HET and tests are already conducted with $N_2$, $O_2$ and Xe as propellants [8]. In Stuttgart, Germany, the IRS is developing an ABEP system based on an intake and an RF Helicon-based plasma thruster (IPT), especially suited to survive long time exposure to chemically aggressive species, like atomic oxygen in VLEO with its electrodeless thruster design [4]. In Russia, TsAGI proposed a concept of ABEP for maintaining a low-orbit spacecraft [9]. In USA, Shabshelowitz proposed and conducted test on two concepts: Radio frequency Plasma Thruster (RPT) and Helicon Hall thruster (HHT) for application of ABEP system [10]. BUSEK, an American company, proposed a concept of Martian Atmosphere-Breathing Hall-Effect Thruster (MABHET) which is an ABEP system designed for the Martian atmosphere [11]. In Japan, the Atmosphere Breathing Ion engine (ABIE) is a concept developed by the JAXA that consists of an intake, a discharge chamber, grids, and a neutralizer with an Electron Cyclotron Resonance (ECR) ion thruster proposing its feasibility in the altitude range of 140 km to 200 km around Earth with the system tested in pulsed operation [12].

While the above literature gives us an idea on different thruster combinations currently being proposed and tested for an ABEP system, in this paper, novel mission scenarios for the application of ABEP system in different environments are presented and analyzed. Beginning with the EO case, namely maintenance of a circular orbit around the Earth, the complexity of the mission scenarios is consistently increased. The five different scenarios analyzed in the paper are:

1. Circular orbit maintenance around Earth.
2. Orbit raising and constant rate de-orbiting around Earth.
3. Elliptical orbit maintenance around Earth.
4. Circular orbit maintenance around Mars.
5. Space tug and refuelling missions in Earth and Mars orbits.

The main focus of the analysis is the determination of the required power levels and system efficiencies (intake efficiency $\eta_c$ and thruster efficiency $\eta_t$) for each mission scenario from which its general feasibility is evaluated. In the following, the methodology of all mission scenarios is shortly described:

**A. Circular orbit maintenance around Earth**

For circular orbit maintenance, the atmospheric drag experienced by the spacecraft needs to be fully compensated by the ABEP system at any time. Therefore, depending on the theoretically available mass flow rate $\dot{m}$, the required exhaust velocity $v_e$ as well as the corresponding required power $P_{req}$ for full drag compensation (FDC) are determined. Conversely, based on an estimated available power, the required parameters of an ABEP system, namely the intake efficiency $\eta_c$ and the thruster efficiency $\eta_t$, to enable FDC can be determined. In terms of available power, the end-of-life power of GOCE [13] is used to establish a realistic baseline scenario.

**B. Orbit raising and constant rate de-orbiting around Earth**

Orbit raising and constant rate de-orbiting are realized by in- and decreasing the thrust required for orbit maintenance, respectively. Based on the baseline scenario, the required $\eta_c$ and $\eta_t$ are estimated.

**C. Elliptical orbit maintenance around Earth**

To maintain elliptical orbits, a thrusting arc, during which the change in orbital parameters due to drag is compensated for, is defined. By comparing the resulting orbital parameters with the ideal Keplerian motion, the requirements for orbit maintenance are extracted.

**D. Circular orbit maintenance around Mars**

Orbit maintenance around Mars is performed similar to that around Earth by applying different boundary conditions (i.e., atmospheric properties).

**E. Space tug and refuelling missions in Earth and Mars orbits**

The applicability of ABEP for space tug and refuelling mission scenarios is analyzed. In the space tug scenario, the ABEP based spacecraft carries the payload to its destination orbit by using the propellant collected at low altitudes, deploys it, and then returns back to its working orbit and collects the propellant. In the refuelling scenario, the ABEP-based spacecraft collects the propellant and refuels the target spacecraft that then detaches and proceeds to its destination orbit. Propellant collection can be realized when the applied power exceeds $P_{req}$ for drag compensation and, at the same time, no orbit raising is performed.

**2. System analysis**

As of today, the suitable altitude range for an ABEP application (VLEO) is practically unused [14]. This imparts ABEP propelled spacecraft with several advantages in comparison to conventional satellites. The benefits can be classified into two distinct mission categories: missions around Earth and missions around other celestial bodies. Around Earth, operating at lower altitudes can be advantageous for commercial applications while on other celestial bodies it can help gain deeper understanding of their atmospheric composition and surface topology. An overview of the benefits of VLEO for Earth observation can be found in [2]. Whereas altitude limits for the VLEO range have already been defined, currently there exists none for other celestial bodies. Therefore, in the following, the respective limits (upper and lower) of altitude are exemplarily derived for Mars.

### A. Simulation setup

Throughout the analysis, a generic spacecraft (size and shape comparable to the GOCE spacecraft Figure 2 (left) [15]) is assumed. The respective parameters are shown in Table 1.

Table 1: Spacecraft properties assumed for mission simulation.

| Mass, $m$ | Frontal area, $A_f$ | Drag coefficient, $C_D$ |
|---|---|---|
| **1000 kg** | 1 m² | 3.7 [16] |

The spacecraft is assumed to be equipped with the Enhanced Funnel Design (EFD) intake [17] shown in Figure 2 (right), as well as the RF Helicon-based Plasma Thruster (IPT), both developed at the Institute of Space Systems of the University of Stuttgart [4,6]. The thruster is unique due to its antenna design, it is compatible with different propellant species due to its electrodeless design which increases its lifetime and also eliminates the need for a neutralizer by ejecting a quasi-neutral plasma. For the thruster efficiency $\eta_t$, the maximum value for a Helicon plasma thruster achieved in literature to date of $\eta_t = 0.20$ is used [18]. In the case of the intake efficiency $\eta_c$, a value for the EFD of $\eta_c = 0.43$ is used [17]. In addition, a second value of $\eta_{c,max} = 0.70$ is used to gauge the results at better intake designs.

To simplify the mission analysis, the complete frontal area of the spacecraft is assumed to be used as intake. Thus, the intake area is $A_{in} = A_f = 1$ m² as well.

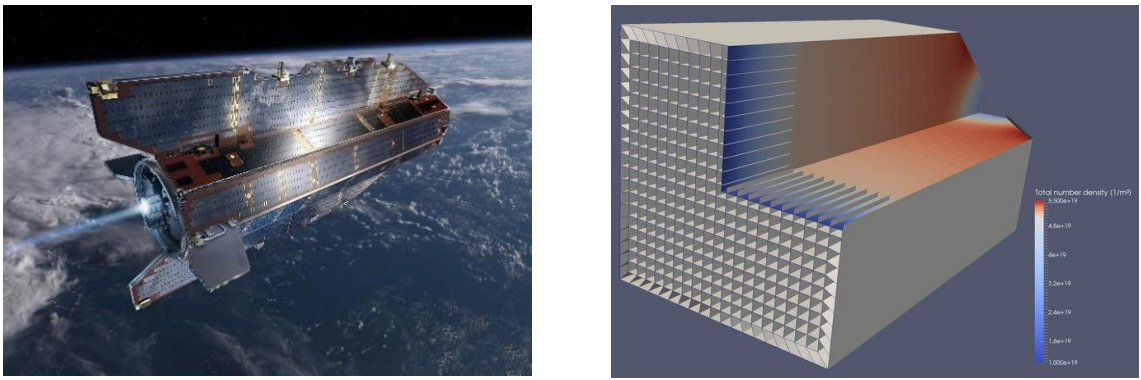

Figure 2: GOCE spacecraft (left) [19] and EFD intake design (right) [20].

The simulations are performed in MATLAB® using the NLRMSISE-00 environmental model [21] for the atmospheric density and composition of Earth and the Mars Climate Database (MCD) v5.3 for Mars [22], respectively. Throughout the analysis, moderate and constant solar and geomagnetic activity settings ($F_{10.7} = 140 \; sfu$, $A_p = 15$) are applied [23].

### B. Definition of the very low orbit regime limits

The feasibility of the ABEP system can only be attained in altitude ranges with a sufficiently dense atmosphere. These regimes can be referred to as very low orbits of the celestial body. The following section determines the upper and lower limits of Earth and Mars, while the same methodology can be applied for any other celestial bodies with an atmosphere.

### A.1. Upper limit

Whereas no standard definition is available yet, the upper limit of VLEO has been defined at 450 km [2]. Below this altitude, the aerodynamic drag causes a conventional spacecraft to decay in less than 5 years (depending on the spacecraft's physical attributes) and therefore necessitates significant changes in conventional spacecraft designs. To define the upper limit for Mars, the idea is to determine the altitude at which the resulting drag force equals the respective value at 450 km around Earth.

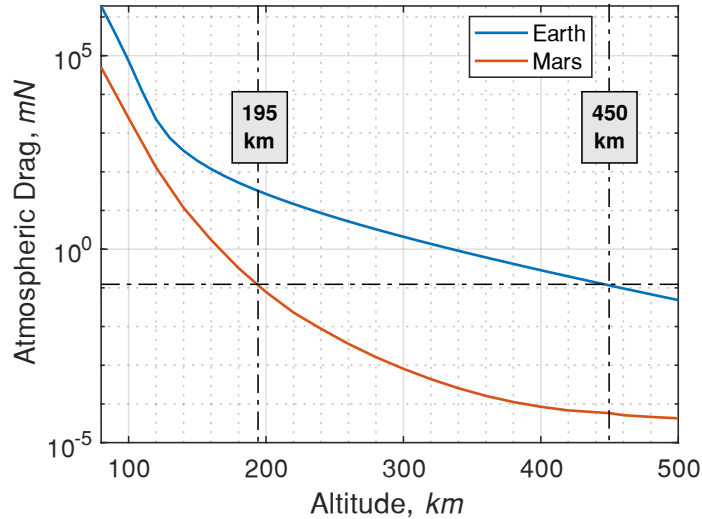

Figure 3: Upper limit for VLEO and VLMO from NRLMSISE-00 and MCD v5.3 model for moderate solar activity with $C_D = 3.7$ and $A_f = 1\ \text{m}^2$ for satellite in a circular orbit.

The norm of the drag force acting on the spacecraft $F_D$ is calculated via the drag equation:

$$F_D = \frac{1}{2} \rho\ C_D\ A_f\ v^2 \qquad (1)$$

Here, $\rho$ is the atmospheric density, $v$ is the spacecraft's orbital velocity, $A_f$ is the frontal area of the spacecraft and $C_D$ is the drag coefficient. Notably, the co-rotation of the atmosphere as well as thermospheric wind effects are not taken into account, which can cause under- or overprediction of the drag effects. These effects will be considered in the future work.

As can be seen from Figure 3, the drag force acting on a spacecraft at 450 km around Earth is experienced on Mars at an altitude of approximately 195 km. Hence, this value is used as the upper limit for VLMO throughout this paper. However, it should be noted that VLEO and VLMO upper limits does not necessarily give the altitude range up to which an ABEP system can operate as this would depend upon on the thruster employed. The limit of minimum $\dot{m}$ and pressure required for the thruster operation can differ for each thruster and should be obtained experimentally. Depending upon the altitude, the ABEP and the thruster should be designed following an iterative process. Doing so can help in evaluating the upper operational altitude for a given ABEP system. For the considered thruster the ignition limits at least show promising results in terms of the needed minimum density. This is expected and also confirmed by experience where the IRS IPG and IPT could even be operated on the background gas.

### A.2. Lower limit

The lower limits of the VLEO and VLMO are determined by the maximum heat load acting on the spacecraft beyond which it would need additional thermal control devices (e.g., a heat shield). To do so, the resulting heat flux acting on the spacecraft is compared to the upper limit of a reference mission and the resulting steady state spacecraft temperature $T_{sc}$ is determined.

The overall heat flux $Q_{Tot}$ acting on the spacecraft is calculated as a first assumption by considering the spacecraft to be a perfect black body with coefficients of absorption and emission equal to 1 and summing up all heat fluxes experienced by the spacecraft. The respective equation is: [24]

$$Q_{Tot} = Q_{Drag} + Q_{Solar} + Q_{IR-Earth} + Q_{Albedo} \tag{2}$$

Here, $Q_{Drag}$ is the heat flux due to aerodynamic drag, $Q_{Solar}$ is the heat flux due to solar radiation, $Q_{IR-Earth}$ is the heat flux due to infrared radiation of Earth and $Q_{Albedo}$ is the heat flux due to albedo effects. The values are calculated according to [24, 25]. From the total heat flux, emissivity ($\epsilon = 1$ for black body) and Stefan–Boltzmann constant $\sigma$, the resulting steady state temperature of the spacecraft $T_{sc}$ is determined via Eq. (3):

$$T_{sc} = \left(\frac{Q_{Tot}}{\epsilon\,\sigma}\right)^{\frac{1}{4}} \tag{3}$$

Figure 4 shows the resulting steady state temperature of a spacecraft $T_{sc}$ orbiting around Earth and Mars for altitudes below 250 km. In order to define an upper threshold for the heat flux, ESA's Solar Orbiter mission is taken as a reference mission. This consists of a satellite which experienced a total heat flux of $Q_{Tot} = 28000$ W/m² (at 0.23 AU) and used a heat shield to sustain this flux [26]. Around Earth, the same heat flux is experienced by the spacecraft at an altitude of around 110 km where the spacecraft steady state temperature reaches $T_{sc} = 845$ K. At a slightly higher altitude of 120 km where it can experience comparably lower temperature of $T_{sc} = 603$ K, the spacecraft could dissipate the excess heat via active or passive methods such as using heat pipes and radiators, while allowing operational thermal conditions for the subsystems [27]. Thus, 120 km is defined as the lower limit of VLEO. Similarly, around Mars, the lower limit of VLMO is defined by calculating the altitude at which the same steady state temperature of $T_{sc} = 603$ K is observed. Thereby, the lower limit of VLMO is found to be 95 km.

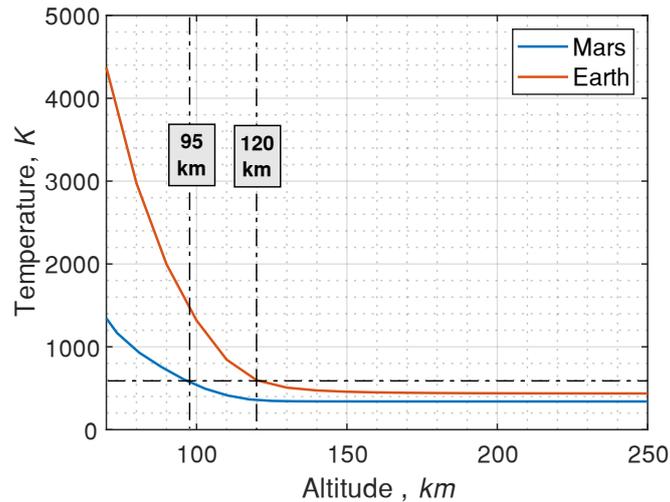

Figure 4: Lower limit for VLEO and VLMO with NRLMSISE-00 and MCD v5.3 for moderate solar activity with $C_D = 3.7$ and $A_f = 1$ m² for satellite in a circular orbit.

### A.3. Overall limits for Earth and Mars

In Table 2, the overall limits for VLEO and VLMO as derived in the previous subchapters and used throughout the rest of this publication are summarized:

Table 2: Upper and lower limits of VLEO and VLMO.

| Altitude | Defining parameter | Earth | Mars |
|---|---|---|---|
| **Lower limit** | Heat load $Q_{Tot}$ | 120 km | 95 km |
| **Upper limit** | Drag force $F_D$ | 450 km | 195 km |

## 3. Analysis

### A. Circular orbit maintenance around Earth

To maintain the altitude at a constant level, the thrust $T$ provided by the ABEP system must compensate the local drag force $F_D$ so that the net force acting on the satellite is nullified:

$$T = F_D \tag{4}$$

The available mass flow rate $\dot{m}$ for the thruster can be calculated via:

$$\dot{m} = \eta_C \, \rho \, A_{in} \, v \tag{5}$$

The required thrust for FDC as well as the available mass flow for a circular, equatorial orbit at 200 km for both different intake efficiencies are shown in Figure 5.

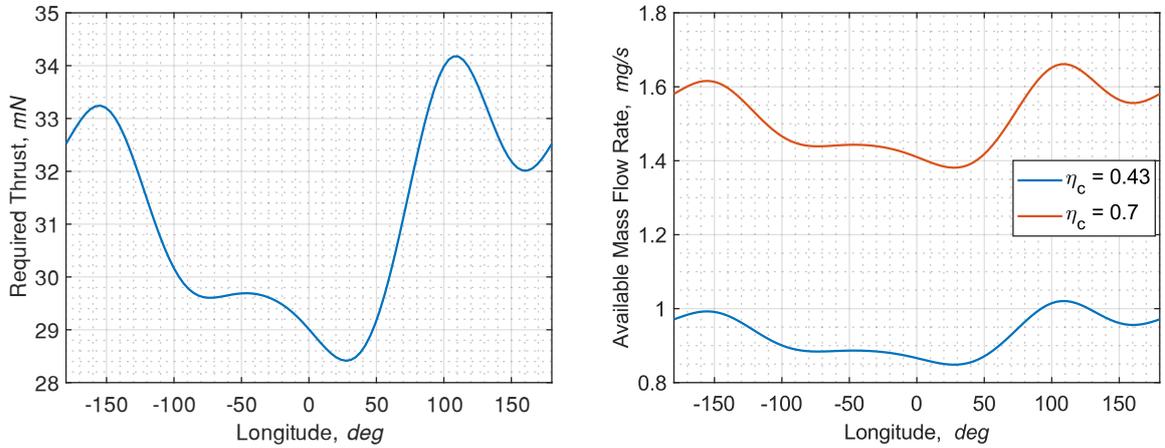

Figure 5: Required thrust (left) and available mass flow rate (right) at 200 km using NRLMSISE-00 model for moderate solar activity with $A_f/A_{in} = 1$ and $C_D = 3.7$.

The thrust provided by the ABEP system can be calculated via spacecraft thrust equation [28]:

$$T = \dot{m} v_e \tag{6}$$

By inserting this expression into Eq. (4) it follows:

$$\dot{m} v_e = \frac{1}{2} \rho \, C_D A_f v^2 \tag{7}$$

This equation can be rearranged for the required exhaust velocity $v_e$ according to:

$$v_e = \frac{v\, C_D}{2\, \eta_c} \frac{A_f}{A_{in}} \quad (8)$$

Note that for the assumed spacecraft geometry $A_f/A_{in} = 1$. Under this assumption, the exhaust velocity for FDC in a circular orbit (constant orbital velocity) is constant. The exhaust velocities required for FDC in a circular orbit at 200 km are listed in Table 3.

Table 3: Exhaust velocities at 200 km required for full drag compensation.

| Intake efficiency, $\eta_c$ | 0.43 | 0.70 |
|---|---|---|
| Exhaust velocity, $v_e$ | 33.49 km/s | 20.57 km/s |

From the definition of the thruster efficiency $\eta_t$:

$$\eta_t = \frac{P_{jet}}{P_{in}} = \frac{\frac{1}{2}\dot{m}v_e^2}{P_{in}} \quad (9)$$

The power $P_{in} = P_{req}$ which is the required power for FDC and can be calculated as:

$$P_{req} = \frac{\dot{m}v_e^2}{2\eta_t} \quad (10)$$

In Eq. (9), $P_{jet}$ is the jet power. The required power for FDC at an altitude of 200 km and an equatorial orbit for both different intake efficiencies is plotted and shown in Figure 6.

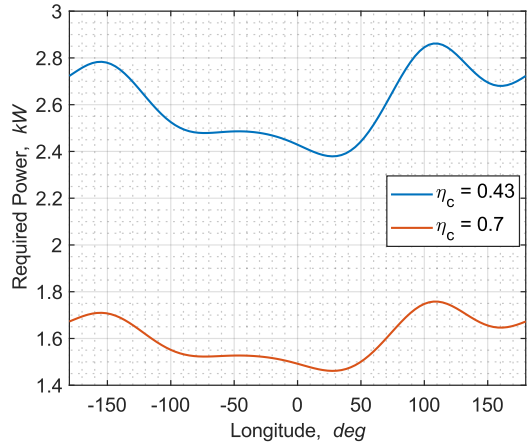

Figure 6: Required power over longitude at 200 km using NRLMSISE-00 model for moderate solar activity for $\eta_t = 0.20$ with $A_f/A_{in} = 1$ and $C_D = 3.7$.

Using the results depicted in Figure 6, the solar array area $A_{sa}$ required to satisfy the peak power requirements at the end-of-life (EOL) is calculated assuming beginning-of-life (BOL) solar cell efficiency of 30.2% and 2.75% degradation per year over a mission lifetime of 2 years [29].

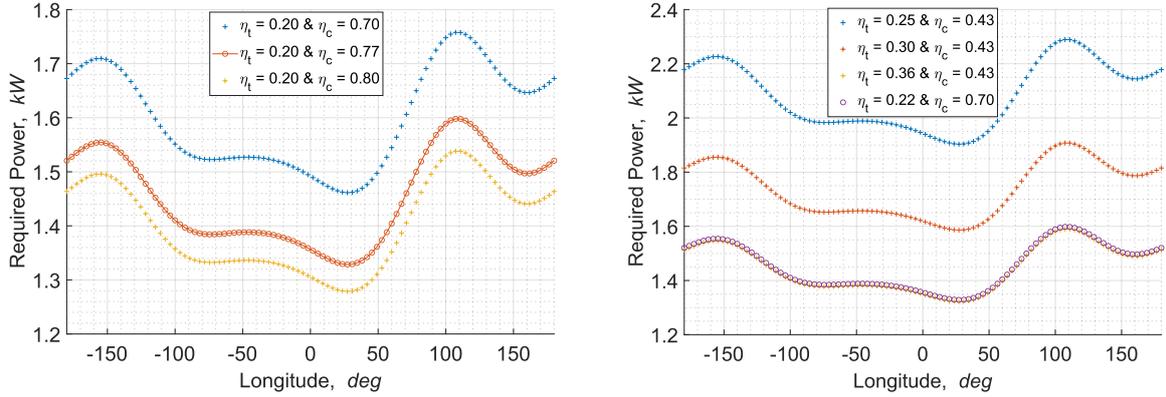

Figure 7: Required power levels for varying intake and thruster efficiencies at 200 km using NRLMSISE-00 model for moderate solar activity with $A_f/A_{in} = 1$ and $C_D = 3.7$.

This results in a required effective solar array area of $A_{sa} = 7.5$ m$^2$ ($\eta_c = 0.43$) and $A_{sa} = 4.67$ m$^2$ ($\eta_c = 0.70$) to ensure the provision of the EOL peak power values of $P_{req} = 2.87$ kW and $P_{req} = 1.75$ kW, respectively, as can be seen in Figure 6. Also, it should be noted that an ABEP based spacecraft should be as aerodynamically optimized to minimize $A_f$ and $C_d$ within a respective mission frame, thereby, the drag due to it. Hence, body-mounted solar arrays are more suitable than the deployable ones. However, this design would require larger area than the effective $A_{sa}$ to attain the required power as depicted in Figure 6. The effective surface area of the solar array $A_{sa}$ can be maintained by making the spacecraft longer while maintaining the same $A_f$. This will lead to an increase in $C_D$, however, its value of 3.7 assumed for the simulation is high enough to get a first order approximation as it has been a common practice to assume a constant $C_D$ of 2.2 for LEO satellites [30].

To assess if $P_{req}$ for the operation of ABEP system are achievable in a realistic mission scenario, the results are compared to the GOCE mission, which operated in a near polar, Sun-synchronous Orbit (SSO) at a mean altitude of 268 km [31]. The spacecraft had comparable specifications as those assumed for the ABEP spacecraft such as similar frontal area, design to minimize aerodynamic drag, and low altitude operation. Considering these similarities, and assuming that the ABEP spacecraft would also have a mission lifetime of at least 2 years, it should be able to produce an EOL power as that of GOCE i.e., $P_{GOCE} = 1.6$ kW, which is considered as the maximum reference power achievable.

From Figure 7, it can be noticed that by varying either of the two parameters i.e., $\eta_c$ and $\eta_t$, individually and keeping the other constant as per the assumptions and imposing the required power of $P_{GOCE} = 1.6$ kW, FDC can be achieved for three different combinations of thruster and intake efficiencies (as shown in Table 4):

Table 4: System parameters for orbit maintenance at 200 km.

| System parameters | Combination 1 | Combination 2 | Combination 3 |
|---|---|---|---|
| $\eta_t$ | 0.20 | 0.22 | 0.36 |
| $\eta_c$ | 0.77 | 0.70 | 0.43 |

The same analysis has been performed for the altitudes of 160 km, 180 km, and 250 km. The results indicate that the thrust required for FDC at 250 km can be achieved with the state-of-the-art efficiencies and an available power of 1.6 kW. At 160 km, the intake and thruster efficiencies must be increased to 0.86 and 0.77, respectively, to achieve FDC. Taking the current benchmarks of the system efficiencies as a basis, these values are highly unlikely to be achieved in the near future. For 180 km, considering an intake efficiency of 0.70, the required power level can be reduced to 1.6 kW by improving the thruster efficiency to 0.43. However, it should be noted that it is not necessary to strictly confine to the given reference power level of 1.6 kW, since by a trade-off between the solar array area and the corresponding drag, the FDC at lower altitudes is attainable.

Also, the frontal area can be reduced to have a trade-off between required $\dot{m}$ and the drag experienced maintaining the reference power level, thereby, leading to an increase in $C_D$, as discussed earlier. The simulation of such a mission has been carried out at an altitude of 180 km by reducing the frontal area to 0.465 m² which is approximately half of the area assumed during the initial simulation, and it is observed that the FDC can be obtained with the assumptions mentioned in the simulation setup.

**B. Orbit raising and constant rate de-orbiting around Earth.**

**Orbit raising**

To perform orbit raising, the total or required thrust ($T_{tot}$) produced by a spacecraft needs to exceed the thrust required for FDC (referred to as $T_D$ in the following) by a certain amount dedicated to raise the orbital altitude $T_{or}$:

$$T_{tot} = T_D + T_{or} \quad (11)$$

To calculate the additional thrust needed for orbit raising ($T_{or}$), the altitude range from the initial (working) orbit to the target orbit is divided into a discrete number of intervals ($n$) and the required thrust for each interval is estimated via the respective equations for a low thrust trajectory [32].

$$T_{or} = m_{sc} \frac{v}{\Delta t} \left[1 - \sqrt{\frac{r_o}{r}}\right] \quad (12)$$

Here, $v$ is the spacecraft velocity (dependent on the altitude), $m_{sc}$ is the spacecraft mass, $\Delta t$ is the transfer time, $r_o$ and $r$ are the initial and target orbital radii, respectively. In the following, the orbit raising maneuver from a reference altitude of 200 km to a target altitude of 250 km is analyzed. To ensure that the power levels are within the limits of $P_{GOCE}$, any of the combinations of intake and thruster efficiencies from Table 4 can be used. However, for this chapter, Combination 3 has been employed for the simulations of both orbit raising and de-orbiting maneuvers. The altitude range is subdivided into $n = 100$ intervals with an additional thrust of 2 mN along with thrust required to maintain the orbit that results into a total transfer time of 170 days. The results are depicted in terms of the required thrust and power in Figure 8. As the spacecraft's altitude increases, the amount of thrust required, and therefore, the required input power reduces as a result of the decreasing density and is well within the limit of 1.6 kW.

Assuming that the spacecraft remains in the sunlight region for FDC, a more optimal orbit raising maneuver in terms of time can be envisaged. This is because the required power keeps reducing with increasing altitude while the available power remains constant. In this case, the spacecraft accelerates as it increases altitude by gradually reducing the respective time in each of the 100 intervals ($\Delta t_n$) to maintain the power level below 1.6 kW. By iteration, the minimum transfer period for the satellite between 200 km to 250 km altitudes is found to be approximately 100 days, below which the power required by the satellite would exceed the reference value. This is shown in Figure 9 in which the additional thrust is increased from the initial value of $T_{tot} - T_D = 2$ mN to a maximum value of $T_{tot} - T_D = 8.2$ mN to stay within the defined power limit of 1.6 kW.

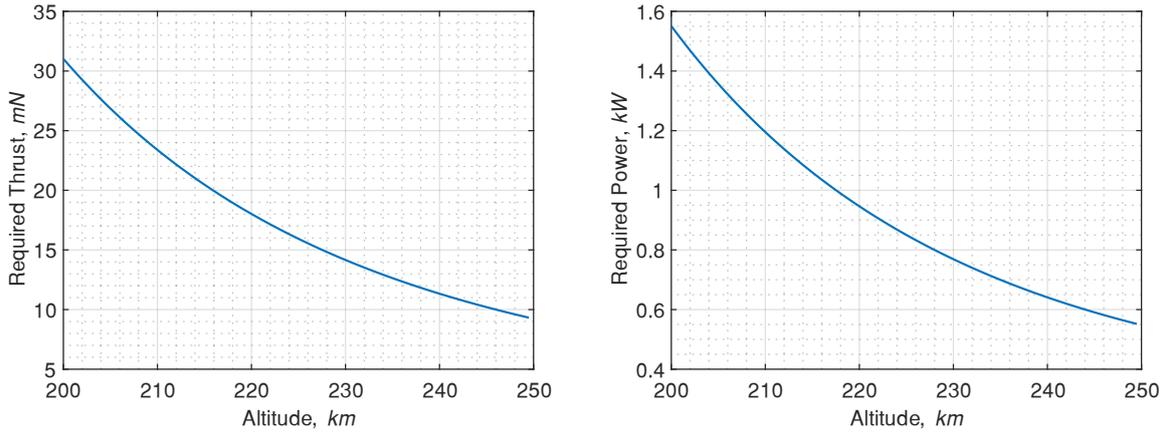

Figure 8: Required thrust (left) and power (right) for constant rate orbit raising with a constant thrust difference of 2 mN using NRLMSISE-00 model for moderate solar activity with $A_f/A_{in} = 1$ and $C_D = 3.7$.

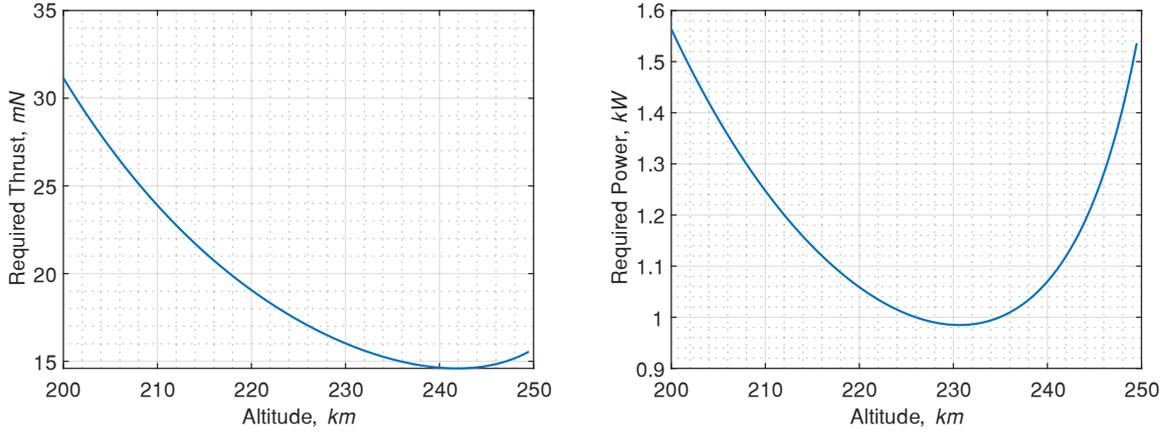

Figure 9: Required thrust (left) and power (right) for accelerated orbit raising with a variable thrust difference of 2 mN to 8.2 mN using NRLMSISE-00 model for moderate solar activity with $A_f/A_{in} = 1$ and $C_D = 3.7$.

**De-orbiting**

To achieve constant rate de-orbiting, the total or required thrust $T_{tot}$ by the spacecraft must be less by a constant value of $T_{de-or}$ than the thrust needed for FDC which is $T_D$.

$$T_{tot} = T_D - T_{de-or} \tag{13}$$

Figure 10 shows the thrust and power levels needed for de-orbiting from 250 km to 180 km at a constant rate. In this case, the resulting thrust is adjusted so that $T_{de-or} = 2$ mN, and a duration of $\Delta t = 238$ days for the maneuver is observed.

To optimize the amount of power required at lower altitudes and the transfer time of the maneuver, the maximum rate at which the spacecraft can de-orbit with a constant rate can be calculated by considering a thruster-off scenario at 250 km. To keep the decay rate constant, the thrust must be gradually increased until the spacecraft reaches its lower target altitude. For the analyzed case, the instant thrust deficit after switching off the thruster is found to be 7.2 mN at 250 km, which is then kept constant throughout rest of the maneuver(Figure 11). Doing so, the transfer time can be reduced to $\Delta t = 66$ days. Notably, if the constant rate orbit decay is not mandatory, the transfer time can be reduced even further by either maintaining the thruster off throughout the mission or having differential deceleration based on the requirements of the maneuver.

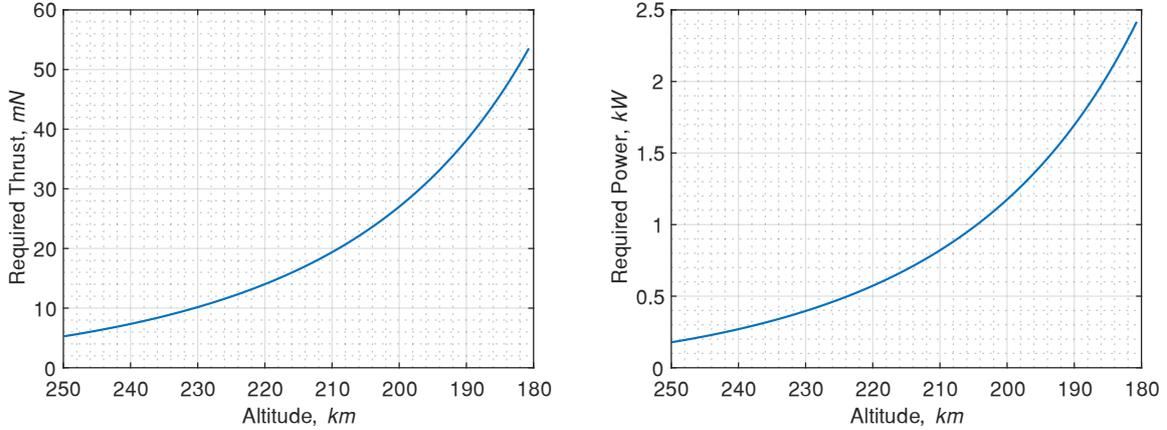

Figure 10: Required thrust (left) and power (right) for a constant rate de-orbiting with a constant thrust difference of 2 mN using NRLMSISE-00 model for moderate solar activity with $A_f/A_{in} = 1$ and $C_D = 3.7$.

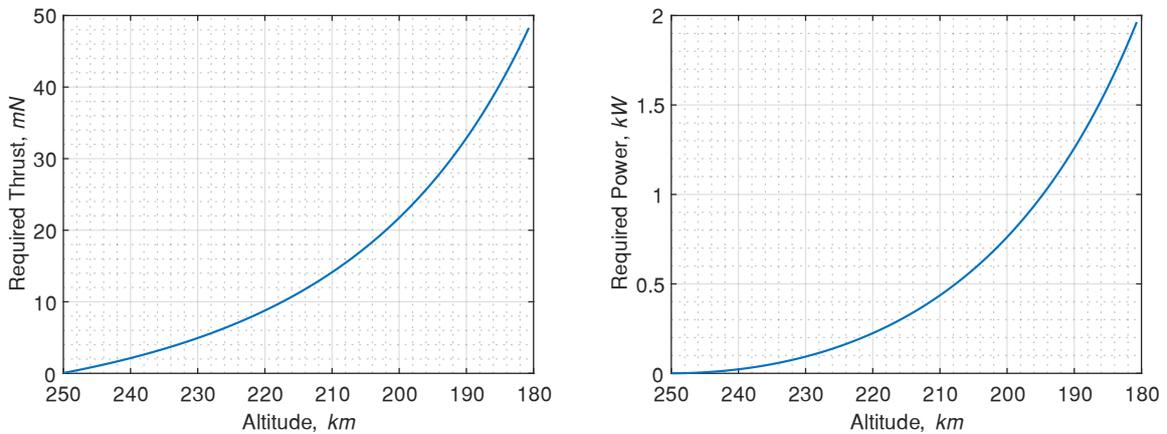

Figure 11: Required thrust (left) and power (right) for a constant rate de-orbiting with a constant thrust difference of 7.2 mN using NRLMSISE-00 model for moderate solar activity with $A_f/A_{in} = 1$ and $C_D = 3.7$.

## C.  Elliptical orbit maintenance around Earth

The Keplerian elements semi-major axis $a$, eccentricity $e$, inclination $i$, right ascension of the ascending node (RAAN) $\Omega$, argument of perigee $\omega$ and true anomaly $\nu$ are used to define the orbit of interest [32]. The original orbit can be maintained by ensuring that these parameters remain same over the lifetime of the satellite. Throughout the analysis, only the effect of aerodynamic drag as a perturbative force is considered. This is a function of satellite's shape, size, mass, material, and orientation with respect to the atmosphere. Moreover, the intensity of solar activity and solar geomagnetic conditions affect the physical characteristics of the atmospheric composition that in turn influences the drag experienced by the satellite. As drag secularly affects only $a$ and $e$, the maintenance task is to keep these two parameters, and thus the size and shape of the orbit but not its orientation, constant. The rate of change of both parameters due to aerodynamic drag with respect to time are as shown in Eq. (14) and Eq. (15) [32]:

$$\frac{da}{dt} = \frac{2F_D a^{3/2}}{m\sqrt{\mu(1-e^2)}}[e \sin\nu \sin\gamma + (1 + e\cos\nu)\cos\gamma] \tag{14}$$

$$\frac{de}{dt} = -\frac{F_D}{m}\sqrt{\frac{a(1-e^2)}{\mu}}[\sin\nu \sin\gamma + (\cos\nu + \cos E)\cos\gamma] \tag{15}$$

Here, $\gamma$ is the flight-path angle, $E$ is the eccentric anomaly, $\mu$ is the gravitational parameter of the Earth and $F_D$ is aerodynamic drag specified in Eq. (1). The rate of change can also be expressed with respect to true anomaly $\nu$ as [32]:

$$\frac{da}{d\nu} = \frac{-\rho(\nu)A_f C_D}{m} a^2 \frac{(1 + e^2 + 2e\cos\nu)^{3/2}}{(1 + e\cos\nu)^2} \tag{16}$$

$$\frac{de}{d\nu} = \frac{-\rho(\nu)A_f C_D}{m} a^2 (1-e^2) \frac{(1 + e^2 + 2e\cos\nu)^{1/2}}{(1 + e\cos\nu)^2}(e + \cos\nu) \tag{17}$$

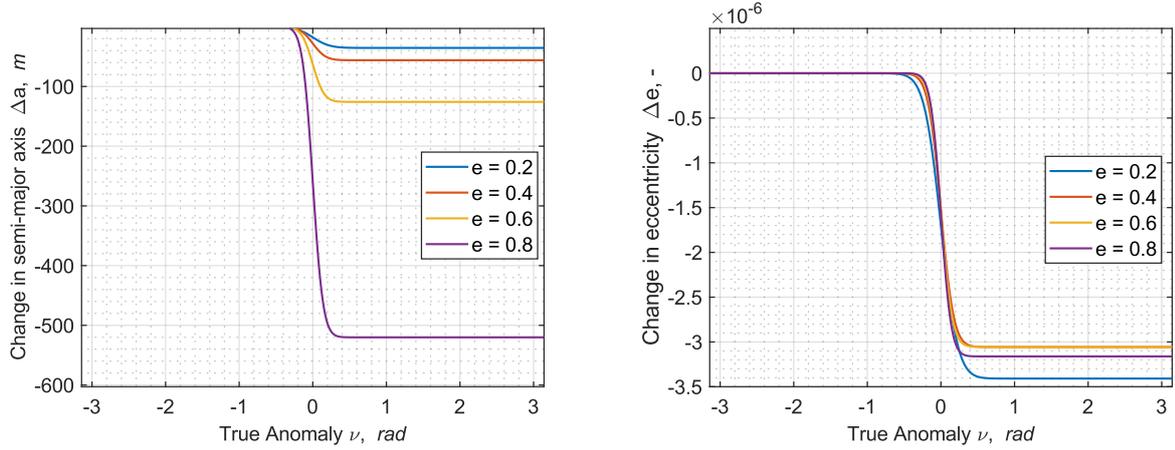

Figure 12: Change in semi-major axis (a) and eccentricity (e) at 200 km perigee altitude for different orbital eccentricities using NRLMSISE-00 model for moderate solar activity with $A_f = 1$ m$^2$ and $C_D = 3.7$.

Figure 12 shows the change in $a$ and $e$ considering the generic spacecraft defined in chapter 2.A. and at a perigee altitude of 200 km over one complete orbit for different values of $e$. It can be seen that for the same perigee altitude, the magnitude of $\Delta a$ and $\Delta e$ increases with $e$.

The thrusting arc required to maintain the original size and shape of the elliptical orbit with perigee altitude $h_p$ is plotted in Figure 13. The results show that this is considerably shorter compared to a circular orbit case with radius $h_p$, which is constantly facing the highest perturbation due to drag. However, due to its increased velocity at perigee, the peak drag force magnitude experienced throughout the orbit is significantly increased for the elliptical orbit. As a result, higher thrust levels are required as shown in Table 5 where it lists the average thrust required for elliptical orbit maintenance is also presented. These values represent the average drag experienced by the spacecraft in the VLEO regime, that is below an altitude of 450 km. Correspondingly, the average power required to maintain the elliptical orbit is lower than that for the circular orbit due to its shorter thrusting period, see Table 5. However, in this case, the peak power is higher as well, see Figure 14 and Table 6.

Table 5: Average and peak thrust levels at 200 km perigee altitude.

| Eccentricity, - | 0 | 0.2 | 0.4 | 0.6 | 0.8 |
|---|---|---|---|---|---|
| Peak thrust, mN | 29.00 | 34.81 | 40.61 | 46.41 | 52.22 |
| Average Thrust, mN | 29.00 | 13.83 | 15.79 | 17.72 | 18.31 |

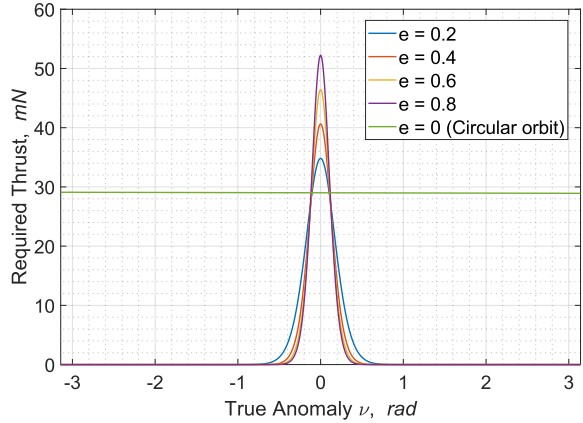

Figure 13: Required thrust for a perigee altitude of 200 km with varying eccentricities using NRLMSISE-00 model for moderate solar activity with $A_f/A_{in} = 1$ and $C_D = 3.7$.

In terms of operations, the electrical power required to cover the peak power levels could be stored during non-thrusting periods. In the end, the total storable power determines whether orbit maintenance is possible for a given orbit. The average and peak power required for elliptical orbits of different eccentricities can be seen in Table 6.

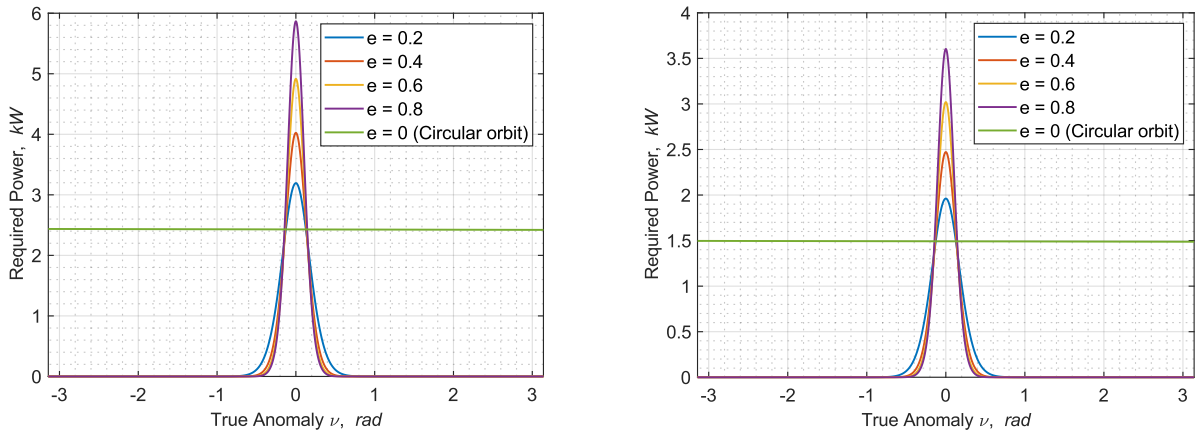

Figure 14: Peak power for a perigee altitude of 200 km with varying eccentricities for $\eta_c = 0.43$ and $\eta_c = 0.70$, $\eta_t = 0.20$ using NRLMSISE-00 model for moderate solar activity with $A_f/A_{in} = 1$ and $C_D = 3.7$.

Table 6: Average and peak power required at 200 km perigee altitude.

| Eccentricity, - | 0 | 0.2 | 0.4 | 0.6 | 0.8 |
|---|---|---|---|---|---|
| Peak power, kW | 2.45 | 3.19 | 4.02 | 4.92 | 5.87 |
| Average power, kW | 2.45 | 1.07 | 1.36 | 1.67 | 1.99 |

Table 7: Minimum perigee altitude for full drag compensation.

| Eccentricity, - | $\eta_c = 0.43$ | $\eta_c = 0.70$ |
|---|---|---|
|  | **Perigee (km)** | **Perigee (km)** |
| 0.2 | 223 | 206 |
| 0.4 | 231 | 214 |
| 0.6 | 239 | 221 |
| 0.8 | 246 | 227 |

With respect to the reference power level of 1.6 kW for an Earth-based satellite, the perigee altitudes above which orbit maintenance of an elliptic orbit is feasible for intake efficiencies of $\eta_c = 0.43$ and $\eta_c = 0.70$ are stated in Table 7.

### D. Circular orbit maintenance around Mars

This section evaluates the feasibility of using an ABEP system in the Martian atmosphere and additionally assessing its competitiveness to conventional EP systems. Two conventional EP systems, used on-board DAWN and GOCE missions, are considered and their specifications listed in Table 8.

Table 8: Sample propulsion systems.

| Propulsion System | Max. Thrust, $T$ (mN) | Max. Power, $P$ (kW) | Max. Mass Flow, $\dot{m}$ (mg/s) |
|---|---|---|---|
| DAWN [33] | 90 | 2.5 | 3.25 |
| GOCE [34] | 50 | 1.17 | 1.58 |

In Figure 15, the maximum achievable thrust of the two systems is plotted along with the respective drag magnitude in a Martian orbit over altitude. From this, the minimum altitude on Mars where FDC can be achieved can be derived. In both cases, frontal areas of $A_f = 1$ m² and the MCD v5.3 atmosphere model using moderate solar activities were considered. For the DAWN propulsion system, the minimum achievable altitude is $h_{min} = 123$ km and for GOCE $h_{min} = 128$ km. Further, from their designed $\dot{m}$ values the maximum propellant required for a satellite lifetime of 2 years can be derived. The respective value, in both cases referring to Xenon, is listed in Table 9.

Table 9: Propellant required for FDC with continuous thrust for a lifetime of two years at $h_{min}$.

| Propulsion System | Required Propellant (kg) |
|---|---|
| DAWN | 204 |
| GOCE | 100 |

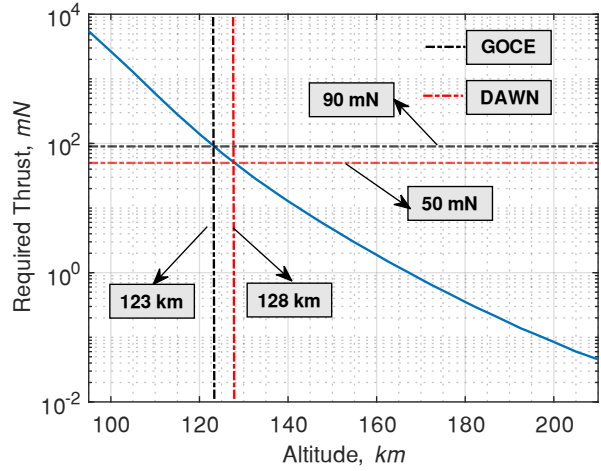

Figure 15: Thrust comparison of different propulsion systems and expected drag (required thrust) using the MCD v5.3 model for moderate solar activity for spacecraft in a circular orbit with $A_f = 1 \text{ m}^2$ and $C_D = 3.7$.

In the next step, the performance of the ABEP system is compared to that of aforementioned propulsion systems. For doing so, the required power for FDC is plotted over altitude. For the Mars based scenario, the reference power ($P_{BUSEK}$) of 1 kW as to the BUSEK concept [35] is considered, since the solar flux at Mars is lower than on Earth, to avoid an overly increased solar array area.

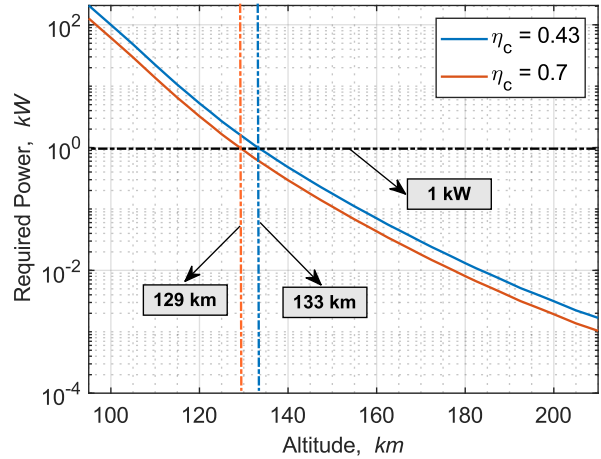

Figure 16: Required power for FDC calculated using the MCD v5.3 model with for moderate solar activity levels and a spacecraft in a circular orbit with $A_f = 1 \text{ m}^2$ and $C_D = 3.7$.

The results plotted in Figure 16 indicate that the ABEP system can achieve FDC at 133 km and 129 km for $\eta_c = 0.43$ and $\eta_c = 0.70$, respectively, and $\eta_t = 0.20$. Comparing this to the required power level and on-board propellant of DAWN and GOCE as seen in Table 8 and Table 9, ABEP proves to be a better option than the conventional EP systems at these lower altitudes of 133 km and 129 km, as it achieves FDC with a power level of 1 kW, while simultaneously mitigating the need of carrying any on-board propellant. As the altitude is increased, the EP devices would need comparatively less propellant and power than that mentioned in Table 8 and Table 9 to stay in orbit for a given lifetime. However, ABEP will still be competitive as the $P_{req}$ for the system reduces with increasing altitude and still having the advantage of longer lifetimes (theoretically dependent only on component durability) than these other EP systems. However, the limiting factor for ABEP would be the available minimum mass flow rate required for its operation. Assessing these trade-offs in closer detail is left for future work.

## E. Space tug and refuelling missions in Earth and Mars orbits

Hereby, the application of ABEP to a spacecraft to extend its potential from being self-sustaining only to additionally collecting and saving propellant for more complex mission scenarios is assessed. Here, two different mission architectures are proposed: 1) Space Tug missions and 2) Refuelling missions. A schematic of both mission architectures is shown in Figure 17. In both cases, the ABEP system collects atmospheric particles and directly stores them. The excess in particles is collected and stored in dedicated tanks. After collecting sufficient amount of propellant, the ABEP equipped spacecraft can either carry a payload to the target orbit (Space Tug architecture) or refuels a target satellite to enable it to reach its target orbit (Refuelling architecture). This paper, however, only assesses the collectable propellant mass for specified conditions and system parameters.

The amount of collectable propellant can be calculated via the following approach, which starts with the definition of the thruster efficiency which is given by Eq. (18):

$$\eta_t = \frac{1}{2}\frac{\dot{m}v_e^2}{P_{in}} \quad (18)$$

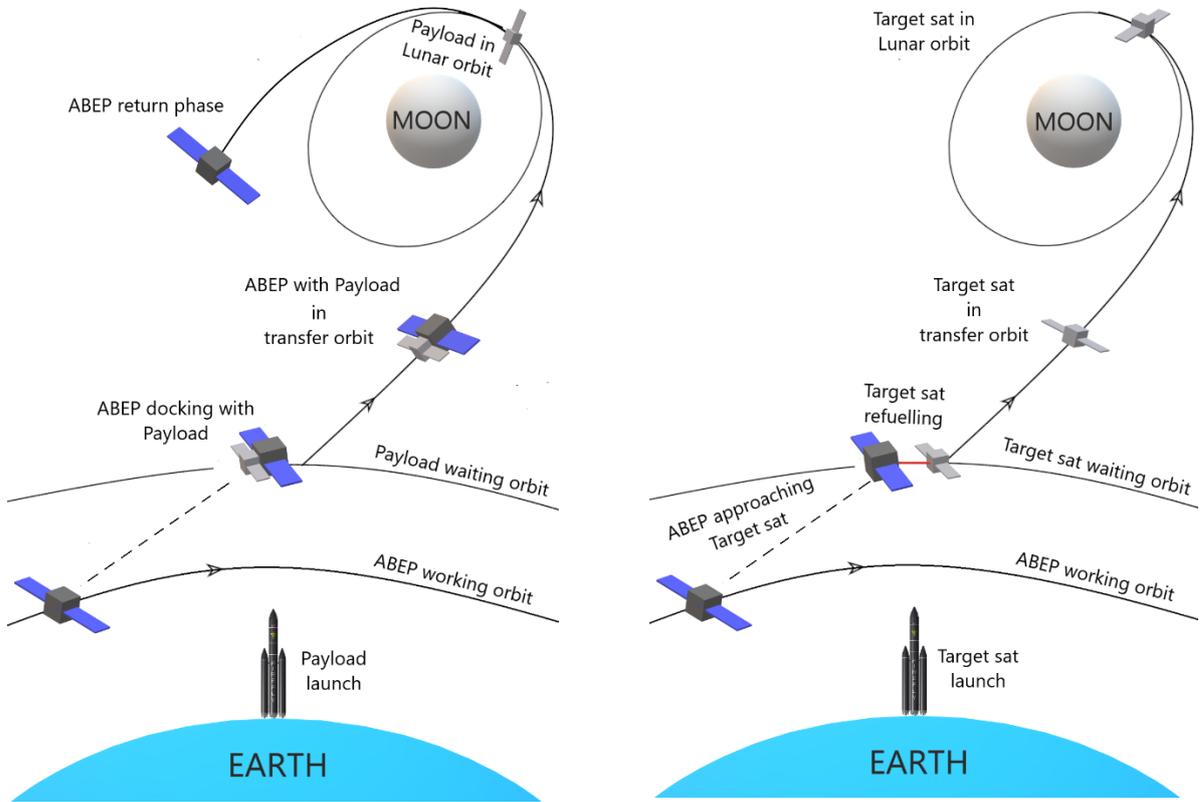

Figure 17: Mission architectures. Space Tug (left) and Refuelling (right).

For FDC, $T = F_D = \dot{m}v_e$ holds, where the total $\dot{m}$ is assumed to equal the incoming ($\dot{m}_{in}$) and outgoing ($\dot{m}_{out}$) mass flow rates i.e., $\dot{m} = \dot{m}_{in} = \dot{m}_{out}$. Rearranging Eq. 18, the power required to achieve FDC $P_{in,FDC}$ can be calculated as in Eq. (19).

$$P_{in,FDC} = \frac{1}{2}\frac{F_D^2}{\dot{m}_{in}\eta_t} \quad (19)$$

Vice versa, by providing any power $P > P_{in,FDC}$ for a given thrust, the contribution of $v_e$ to the jet power can be increased so that a reduced $\dot{m}$ is needed. Doing so, the thrust required by the spacecraft for FDC in the orbit can be achieved while utilizing only a fraction of $\dot{m}_{in}$. This can be expressed by introducing an additional parameter ε, the propellant mass utilization ratio:

$$\varepsilon = \frac{\dot{m}_{out}}{\dot{m}_{in}} \qquad (20)$$

Substituting Eq. (20) in Eq. (19) and rearranging for ε leads to:

$$\varepsilon = \frac{1}{2} \frac{F_D{}^2}{\dot{m}_{in} P \eta_t} \qquad (21)$$

Thus, for any power greater than $P_{in,FDC}$ the storable propellant mass flow $\dot{m}_{stored}$ can be expressed as:

$$\dot{m}_{stored} = \dot{m}_{in} - \dot{m}_{out} \qquad (22)$$

$$\dot{m}_{stored} = \dot{m}_{in}(1 - \varepsilon) \qquad (23)$$

**E.1 Earth referenced scenarios.**
Initially, a comparison is made between a circular orbit and an elliptical orbit to evaluate the amount of propellant that can be collected and stored for a given set of orbital parameters for each orbit type. The propellant collected per year for circular and elliptical orbits at altitudes (perigee altitudes in elliptical orbits, radius for circular orbits) from 180 km to 200 km at 5 kW and $\eta_t = 0.20$, are simulated for both $\eta_c = 0.43$ and $\eta_c = 0.70$, and for different intake areas. The results for the latter are shown in Figure 18. Note that the intake area is still assumed to be equal to the frontal area of the satellite. Altitudes below 180 km are not displayed as this would require either carrying additional on-board propellant or larger power requirements.

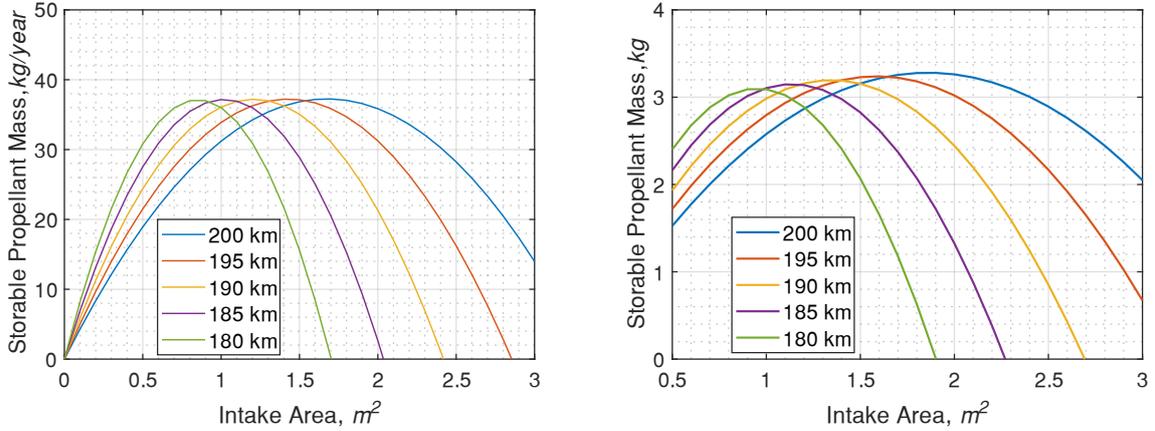

Figure 18: Storable propellant vs intake area for input power of 5 kW in circular and elliptical orbit for $\eta_c = 0.70$ and $\eta_t = 0.20$ using NRLMSISE-00 model for moderate solar activity with $A_f/A_{in} = 1$ and $C_D = 3.7$.

The results depicted in Figure 18 show that for eccentric orbits the storable propellant per year is significantly reduced compared to the circular case. This is due to two factors: the first factor is that the contribution of the velocity to atmospheric drag in an elliptical orbit near the perigee, where the highest velocities occur, is significantly larger than for a circular orbit. Secondly, the satellite in elliptical orbits is assumed to begin propellant collection only during times when it is orbiting inside the VLEO region, i.e., below 450 km altitude, which is only a fraction of the orbital period.

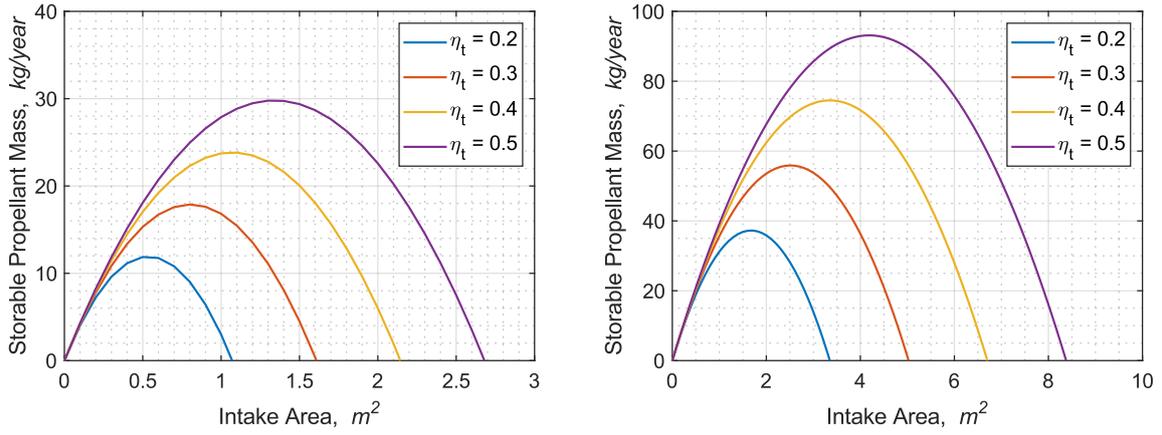

Figure 19: Maximum storable propellant at a given power level at 200 km at 1.6 kW and 5 kW for $\eta_c = 0.70$ using NRLMSISE-00 model for moderate solar activity with $A_f/A_{in} = 1$ and $C_D = 3.7$.

Furthermore, Figure 19 indicates that there is an optimum intake area (also satellite frontal area) for which the storable propellant is maximized. This optimum area for a given power level can be obtained from Eq. (23). This results in the following equation for the intake area size $A_{m_{max}}$ required for maximum collection of propellant for fixed system parameters:

$$A_{m_{max}} = 4\frac{\eta_c \eta_t}{\rho v^3 C_D^2} P \qquad (24)$$

Such insight is highly valuable for future mission planning. In Figure 19 it is shown that the collected propellant can be increased for a given altitude by improving the thruster efficiency $\eta_t$ for the reference level of 1.6 kW and 5 kW.

**E.2 Mars referenced scenarios.**
Around Mars, due to reduced solar flux levels, the same solar array area which can generate 5 kW around Earth can produce around 2.3 kW, which is used as the reference for the simulations following herein.

The storable propellant by mass for different intake areas at different altitudes in VLMO simulated for a power level of 2.3 kW is shown in Figure 20. At 130 km altitude, the maximum mass of the storable propellant is 85 kg/year, similar to that obtained for higher altitudes as shown in Figure 20. Based on Eq. (24), the optimum intake area is found to be about 1.3 m² for 130 km at 2.3 kW. However, the spacecraft needs larger intake areas (peaks of the parabolic curves in Figure 20) as it goes into higher orbits while it still yields the propellant of around 85 kg/year.

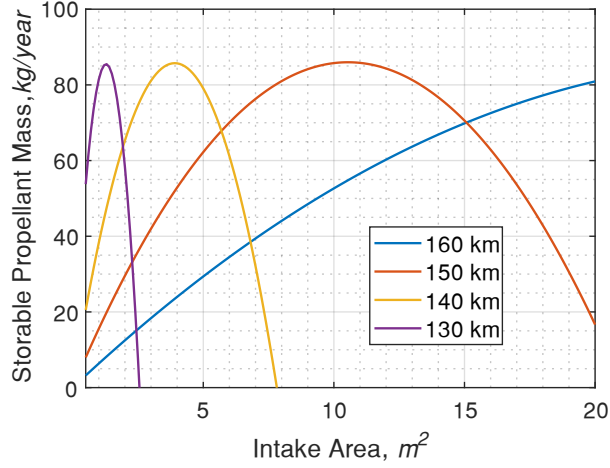

Figure 20: Storable propellant around Mars for varying intake areas for $\eta_c = 0.70$ and $\eta_t = 0.20$ using the MCD v5.3 model for moderate solar activity with $A_f/A_{in} = 1$ and $C_D = 3.7$.

At lower altitudes, instead, the storable propellant is found to be in negative values, which states that the spacecraft would require additional propellant for orbit maintenance rather than collecting it, and therefore, is not considered any further. This could be explained from Eq. (21) to Eq. (23), as ε becomes greater than 1 due to high thrust requirement for FDC at lower altitudes.

The propellant collected by ABEP at 130 km can be increased by improving its $\eta_c$ and $\eta_t$. Since, the simulations have been performed for the assumed $\eta_{c,max}$, Figure 21 shows increase in propellant collection with increasing $\eta_t$.

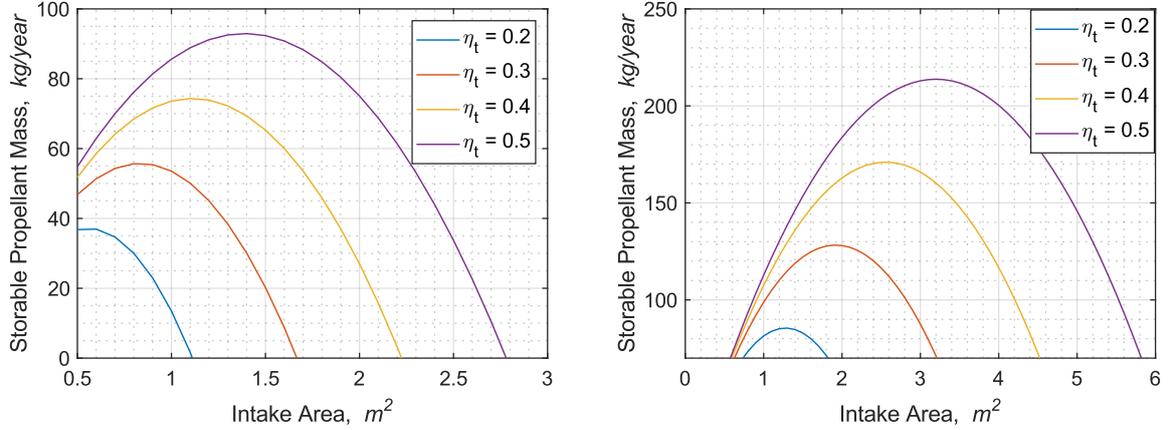

Figure 21: Maximum storable propellant at 130 km for $P_{BUSEK}$ (left) and increased power level of 2.3 kW (right) for $\eta_c = 0.70$ using the MCD v5.3 model for moderate solar activity with $A_f/A_{in} = 1$ and $C_D = 3.7$.

The collected propellant must also be effectively stored for the proposed space tug or refuelling mission. This will require additional systems such as a compression and pumping system [36]. These systems would require power beyond the assumptions presented in the chapter, which was only considered for propulsion. This will analyzed in more detail in the future work of the project.

## 4. Discussion of the results

The following section summarizes the results obtained from the discussed missions, which are broadly categorized into Earth- and Mars-referenced mission scenarios.

**Earth-based scenarios**

For Earth referenced satellites of $A_{in} = A_f = 1$ m$^2$, $C_D = 3.7$ under moderate solar activity, the optimum range for ABEP operation is found to be 180 km to 250 km. The operational requirements for circular orbits are as summarized below:

- For altitudes higher than 200 km and up to 250 km, orbit maintenance with FDC is found feasible for the reference $P_{GOCE}$ of 1.6 kW with $\eta_t = 0.20$, and $\eta_c = 0.43$ and 0.70.
- FDC along with orbit raising and constant-rate de-orbiting maneuvers at 200 km are not achievable with the assumed values of $\eta_t$ and $\eta_c$. Hence, the combinations of $\eta_t$ and $\eta_c$ such as 0.20 and 0.77, 0.22 and 0.70, and 0.36 and 0.43, respectively are proposed to operate at reference $P_{GOCE}$.
- At 180 km, FDC can be achieved at $P_{GOCE}$ by reducing $A_f$ from 1 m$^2$ to 0.465 m$^2$. To simultaneously achieve the same solar array area, more slenderer spacecraft designs can be chosen but that may lead to an increase in $C_D$. However, the value for $C_D$ of 3.7 assumed throughout the analysis is already a conservative estimation.
- Below 180 km, FDC in circular orbit is not feasible as $P_{req}$ exceeds the reference $P_{GOCE}$. In this case, the required power levels vary from 5.5 kW to 12.5 kW for $\eta_c = 0.43$, and 3.5 kW to 7 kW for $\eta_c = 0.70$ as the altitude decreases from 180 km to 160 km.

For an Earth-referenced satellite in elliptical orbits, FDC is achievable ensuring that the perigee altitude above 200 km for the assumed spacecraft characteristics with a reference power equal to $P_{GOCE}$.

- In this case, a minimum perigee altitude $h_{p_{min}} = 206$ km and $h_{p_{min}} = 227$ km for low ($e = 0.2$) and highly ($e = 0.8$) eccentric orbits, respectively, can be maintained.

**Mars-based scenarios**

- ABEP equipped satellite with $A_{in} = A_f = 1$ m$^2$, $C_D = 3.7$ under moderate solar activity can maintain circular orbits from 133 km to 140 km, and 129 km to 135 km for $\eta_c = 0.43$ and $\eta_c = 0.70$, respectively for a thruster efficiency of $\eta_t = 0.20$ within the $P_{BUSEK}$ of 1 kW and assumed system characteristics.
- From 140 km to 195 km (VLMO upper limit), ABEP mitigates the need for on-board propellant with required power levels below 0.5 kW.

**Space Tug and Refuelling**

- Around Earth, the altitude ranges from 180 km to 200 km can have propellant collection of 40 kg/year to 95 kg/year at required power level of 5 kW for $\eta_t = 0.20$ to 0.50 and $\eta_c = 0.70$, below and above which the propellant quantity starts decreasing due to the drag and reduced available $\dot{m}$, respectively.
- Similarly, around Mars, between 130 km and 150 km ABEP can store from 190 kg/year to 475 kg/year at 2.3 kW for $\eta_t = 0.20$ to 0.50 and $\eta_c = 0.70$.

## 5. Conclusion

This paper proposes and analyses several novel ABEP based mission scenarios. Starting from technology demonstration mission in VLEO, more complex mission scenarios are derived and discussed in detail. These include, amongst others, orbit maintenance around Mars as well as refuelling and space tug missions.

The analysis shows that the ABEP system is capable of FDC for continuous orbit maintenance around Earth and Mars within a certain range of altitudes. With the analyzed spacecraft design (comparable in size and shape to the GOCE spacecraft), altitudes from 180-250 km in VLEO and 130-160 km in VLMO are identified as feasible for FDC. To realize drag compensation at even lower altitudes, either the system efficiencies ($\eta_c$ and/or $\eta_t$) or the available power needs to be increased. As for solar powered spacecraft, an increase in the required power requires increased solar array areas (which again leads to increased magnitudes of drag), increasing the system efficiencies is identified as critical for reducing the feasible altitude even further.

Also, more advanced maneuvers like orbit raising and constant rate de-orbiting are shown to be feasible within the aforementioned limits. Therefore, the analysis shows that ABEP systems are promising and flexible propulsion systems in VLEO and VLMO.

With regards to more advanced mission concepts, propellant collection around Mars proves to be a better option compared to around Earth as more propellant can be collected in a given time due to higher $\dot{m}$ available at lower altitudes while having smaller velocity contribution to drag. Thereby, even complex missions such as space tug or refuelling can be made feasible. With the amount of propellant collected around Earth, only microsatellite refuelling is found to be feasible, whereas alternate options should be explored for large cargo transfers.

## 6. Future work

The work presents several applications where ABEP can present an immediate impact. However, each of these applications requires further optimization over the proposed design parameters. Some of the areas that should be tackled are:

- Addition of compression and acceleration stages and their impact on system efficiencies.
- Co-rotation of the planet atmosphere and thermospheric wind effects on drag estimation.
- Hybrid system with ABEP and special spacecraft geometries with/without appendages for orbit control to mitigate orbital decay.
- Efficient storage of the propellant and its effect on power requirements.
- Study around other celestial bodies with atmosphere like Titan, Venus, etc.

The study theoretically illustrates the possibilities of ABEP in a variety of conditions, demonstrating that the system is a promising technology for future space missions that are sustainable due to ISRU and highly productive due to enhanced payload performances owing to its very low altitude operation.

### Acknowledgements

This project has received funding from the European Union's Horizon 2020 research and innovation program under grant agreement No. 737183. This reflects only the author's view, and the European Commission is not responsible for any use that may be made of the information it contains.